\documentclass[10pt, conference, compsocconf]{IEEEtran}
\ifCLASSINFOpdf
\else
\fi
\usepackage{amsopn}
\usepackage{booktabs} 
\usepackage{pgfplots}
\usepackage{subfig}
\usepackage{amsmath,bm}
\usepackage{amsfonts}
\usepackage{tikz}
\usetikzlibrary{patterns}
\usepackage{url} 

\usepackage[outdir=./]{epstopdf}

\usepackage{xspace}

\newcommand{\cpp}{C++\xspace}

\newcommand{\spark}{{Spark}\xspace}

\newcommand{\cocoa}{\textsc{CoCoA}\xspace}

\newcommand{\R}{\mathbb{R}}                      

\newcommand{\vv}{ {\bf v}}

\newcommand{\cv}{ {\bf c}}
\newcommand{\rv}{ {\bf r}}
\newcommand{\alphav}{ {\boldsymbol \alpha}}

\newcommand{\gv}{ {\bf g}}

\newcommand{\bP}{\mathcal{P}}

\newcommand{\bS}{\mathcal{S}}

\newcommand{\A}{($\mathrm{A}$)}
\newcommand{\B}{($\mathrm{B}$)}
\newcommand{\rC}{($\mathrm{C}$)}
\newcommand{\D}{($\mathrm{D}$)}
\newcommand{\E}{($\mathrm{E}$)}

\usepackage{fancyhdr}
\fancypagestyle{firststyle}
{

   \fancyfoot[C]{\footnotesize Copyright \copyright~2017 IEEE. Personal use of this material is permitted. Permission from IEEE must be obtained for all other uses, in any current or future
media, including reprinting/republishing this material for advertising or promotional purposes, creating new collective works, for resale or redistribution to
servers or lists, or reuse of any copyrighted component of this work in other works.
}
}

\begin{document}
%

\title{ Understanding and Optimizing the Performance of Distributed Machine Learning Applications on  Apache Spark}

\author{\IEEEauthorblockN{Celestine D{\"u}nner}
\IEEEauthorblockA{IBM Research \\
Z\"urich, Switzerland\\
cdu@zurich.ibm.com \vspace{-0.4cm}}
\and
\IEEEauthorblockN{Thomas Parnell}
\IEEEauthorblockA{IBM Research \\
Z\"urich, Switzerland\\
tpa@zurich.ibm.com \vspace{-0.4cm}}
\and
\IEEEauthorblockN{Kubilay Atasu}
\IEEEauthorblockA{IBM Research \\
Z\"urich, Switzerland\\
kat@zurich.ibm.com\vspace{-0.4cm}}
\and
\IEEEauthorblockN{Manolis Sifalakis}
\IEEEauthorblockA{IBM Research \\
Z\"urich, Switzerland\\
emm@zurich.ibm.com\vspace{-0.4cm}}
\and
\IEEEauthorblockN{Haralampos Pozidis}
\IEEEauthorblockA{IBM Research \\
Z\"urich, Switzerland\\
hap@zurich.ibm.com\vspace{-0.4cm}}}


%

\maketitle

\begin{abstract}
In this paper we explore the performance limits of Apache Spark for machine learning applications. We begin by analyzing the characteristics of a state-of-the-art distributed machine learning algorithm implemented in Spark and compare it to an equivalent reference implementation using the high performance computing framework MPI. We identify critical bottlenecks of the Spark framework and carefully study their implications on the performance of the algorithm. In order to improve Spark performance we then propose a number of practical techniques to alleviate some of its overheads. However, optimizing computational efficiency and framework related overheads is not the only key to performance -- we demonstrate that in order to get the best performance out of any implementation it is necessary to carefully tune the algorithm to the respective trade-off between computation time and communication latency. The optimal trade-off depends on both the properties of the distributed algorithm as well as infrastructure and framework-related characteristics. 
Finally, we apply these technical and algorithmic optimizations to three different distributed linear machine learning algorithms that have been implemented in Spark. We present results using five large datasets and demonstrate that by using the proposed optimizations, we can achieve a reduction in the performance difference between Spark and MPI from 20x to 2x.
\end{abstract}
%
%

%
\IEEEpeerreviewmaketitle

\thispagestyle{firststyle}


\section{Introduction}
\label{sec:intro}
Machine learning techniques provide consumers, researchers and businesses with valuable insight. The rapid proliferation of machine learning in 
these communities has been driven both by the increased availability of powerful computing resources as well as the large amounts of data that are being 
generated, processed and collected by our society on a daily basis. 
While there exist many small and medium-scale problems that can 
be easily solved using a modern laptop, there also exist datasets that 
simply do not fit inside the memory capacity of a single machine. In order to solve such problems, one must turn to distributed implementations of machine learning: algorithms 
 that run on a cluster of machines that communicate over a network interface.  There are two main challenges that arise when scaling out machine learning to tackle large-scale problems.
The first challenge relates to algorithm design: in order to learn in a distributed environment one must determine how the training data should 
be partitioned across the worker nodes, how the computations should be assigned to each worker and how the workers should communicate with one another 
in order to achieve global convergence. The second challenge relates to implementation and accessibility. Well-established high performance computing 
frameworks such as Open MPI provide rich primitives and abstractions that allow flexibility when implementing algorithms across distributed 
computing resources. While such frameworks enable high performance applications, they require relatively low-level development skills, making them 
inaccessible to many. In contrast, more modern frameworks such as Hadoop and \spark adhere to well-defined distributed 
programming paradigms, provide fault tolerance and offer a powerful set of APIs for many widely-used programming languages. While these abstractions certainly make distributed computing more accessible to developers, they come with poorly understood overheads associated with communication and data management which can severely affect performance.

In this work we aim to quantify and understand the different characteristics of \spark- and MPI-based implementations and, in particular, their implications on the performance of distributed machine learning workloads. Our goal is to provide guidance to developers and researchers regarding the best way to implement distributed machine learning applications.
Therefore we will proceed as follows:

\begin{enumerate}
\item We analyze the performance of a distributed machine learning algorithm implemented from scratch in both \spark and MPI.  We clearly decouple framework-related overheads from computational time in order to study Spark overheads in a language agnostic manner. \vspace{0.01cm}\vspace{-0.3cm} 
\item We propose two techniques for extending the functionality of \spark specifically designed to improve the performance of machine learning workloads. We demonstrate that these techniques, combined with \cpp acceleration of the local solver, provide over an order of magnitude improvement in performance.  \vspace{0.15cm}
\item We demonstrate that, in order to achieve optimal performance using either framework as well as the proposed extensions, it is crucial to carefully tune the algorithm to the communication and computation characteristics of the specific framework being used. 
\vspace{0.15cm}
\item Finally, we study the effect of the proposed \spark optimizations across five large datasets and apply them to  three different distributed machine learning algorithms. We show that  the performance of \spark can be improved by close to $10\times$ in all cases.
 \end{enumerate}

\section{Distributed Machine Learning}
In distributed learning we wish to learn a best-fit classification or regression model from the given training data, where every machine only has access to its own part of the data and some shared information that is periodically exchanged over the network. 
The necessity of this periodic exchange is what makes machine learning problems challenging in a distributed setting and distinguishes them from naively parallelizable workloads.
The reason is that the convergence of machine learning algorithms typically  depends strongly on how often information is exchanged between workers, while sending information over the network is usually very expensive relative to computation.
This has driven a significant effort in recent years to develop novel methods enabling communication-efficient distributed training of machine learning models. 
In what follows we will focus on the class of synchronous learning algorithms.
Such algorithms are more suitable for the purposes of this study than their asynchronous counterparts \cite{Li2014,  multiverso, recht2011hogwild} since they can be naturally expressed as a sequence of  \textit{map} and \textit{reduce} operations and can thus be implemented using the \spark framework.

\subsection{Algorithmic Framework}
\label{sec:alg}
We consider algorithms that are designed to solve 
regularized loss minimization problems of the form 
\begin{equation}
\min_{\alphav\in\R^n} \ell(A\alphav) +r(\alphav),
\label{eq:obj}
\end{equation}
where $A \in \R^{m\times n}$ denotes the data matrix consisting of $m$ training samples and $n$ features.
The model is parameterized by the vector $\alphav\in \R^n$ and $\ell:\R^m\rightarrow \R$ and $r:\R^n\rightarrow \R$ are convex loss and regularization functions respectively.
 The data matrix $A$ contains the training samples $\{\rv_i^\top\}_{i=1}^m$ in its rows and hence the features  $\{\cv_i\}_{i=1}^n$ in its columns.
 The matrix can be partitioned column-wise according to the partition $\{\bP_k\}_{k=1}^K$ such that features $\{\cv_i\}_{i\in \bP_k}$ reside on worker $k$ where $n_k:=|\bP_k|$ denotes the size of the partition.
 Alternatively, the matrix can be partitioned row-wise according to the partition $\{\hat{\bP}_k\}_{k=1}^K$, such that the samples $\{\rv_i\}_{i\in \hat{ \bP}_k}$ reside on worker $k$, where  $m_k:=|\hat{\bP}_k|$.

\begin{figure}[t]
\centering
 \includegraphics[width=8cm]{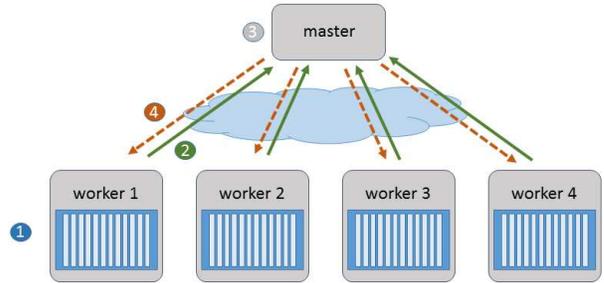} 
\caption{Four-stage algorithmic pattern for synchronous distributed learning algorithms. Arrows indicate the synchronous communication per round.
}
\vspace{-0.7cm}
\label{fig:cluster} 
\end{figure}

The model is then learned in an iterative fashion  as illustrated in Figure~\ref{fig:cluster}:
During one round of the algorithm, a certain amount of computation is performed on each worker (1) 
and the results of this work are communicated back to the master (2). Once the master has received the results from all workers it performs an aggregation step (3) to update some global representation of the model. This information is then broadcast to all workers (4) and the next round can begin.
Let us introduce the hyperparameter $H$ which quantifies the number of local data vectors that are processed on every worker during step (1). This will prove to be a useful tuning parameter allowing users to optimally adapt the algorithm to a given system. In this paper we will cover three prominent algorithms that adhere to this strategy in more detail:

\paragraph{CoCoA} We have implemented the \cocoa algorithm as described in \cite{CoCoA16}. 
The data is partitioned feature-wise across the different workers and the regularization term in \eqref{eq:obj} is assumed to be separable over the coordinates of $\alphav$. 
Given this partitioning, every worker node repeatedly works on an approximation of \eqref{eq:obj} based on its locally available data.
The \cocoa framework is flexible in the sense that any algorithm can be used to solve the local sub-problem. The accuracy to which each local subproblem is solved is directly tied to the choice of $H$. 
For more detail about the algorithm framework and sub-problem formulations, we refer the reader to \cite{CoCoA16}.
In our implementation we use stochastic coordinate descent (SCD) as the local solver, where every node works on its dedicated coordinates of $\alphav$. 
Hence, in every round, each worker performs $H$ steps of SCD after which it communicates to the master a single $m$-dimensional vector:
$\Delta \vv_k:=A\, \Delta \alphav_{[k]},$
where $\Delta \alphav_{[k]}$ denotes the update computed by worker $k$ to its local coordinate vector during the previous $H$ coordinate steps. $\Delta \vv\in \R^m$ is a dense vector, encoding the information about the current state of $ \alphav_{[k]}$, where $\alphav_{[k]}$ itself can be kept local. 
The master node then aggregates these updates and determines the global update $\vv^{(t+1)}=\vv^{(t)} + \gamma\sum_k \Delta \vv_k$ which is then broadcast to all workers and all local models are synchronized. We use $\gamma=1$ for our implementations of \cocoa and tune remaining algorithmic parameters accordingly.

\paragraph{Distributed mini-batch SCD} This algorithm differs from \cocoa only in how the local updates $\Delta \alphav_{[k]}$ are computed. While in \cocoa local SCD updates are immediately applied, for mini-batch SCD, every worker computes the gradient $g_i$ for a subset $\bS_k\subset \bP_k$ of size $H$ of its local coordinates and then determines the update as $\Delta \vv_{k} =- \gamma \sum_{i\in \bS_k} g_i \cv_i$, where $\gamma\in\R^+$ is the stepsize.
The master node  aggregates these updates, updates $\vv$ and broadcasts this vector back to the workers.
\paragraph{Distributed mini-batch SGD} In contrast to the former two algorithms, mini-batch stochastic gradient descent (SGD) requires separability of the loss term in \eqref{eq:obj} over the samples and assumes the data to be distributed row-wise. Then, in every round, each worker  computes a gradient over a subset $\bS_k\subset \{\rv_i\}_{i\in \hat{ \bP}_k}$ of size $H$ of its local samples. These  gradients are then  communicated to the master node which aggregates them to compute an approximation of the global gradient $\tilde \gv\in \R^{n}$ and perform a gradient step on $\alphav$: $\alphav^+ := \alphav-\gamma \tilde \gv$ before the new parameter vector is broadcasted to the workers. 
\vspace{-0.1cm}

\subsection{Performance Model}
\label{sec:model}
After introducing the hyper-parameter $H$, 
 the execution time of a distributed algorithm can be modelled as follows:
Let us denote $N_\epsilon(H)$ the number of rounds needed to achieve a suboptimality of $\epsilon$ given $H$ and $T_\epsilon(H)$ the corresponding execution time. Then, we can write
\begin{equation}
\label{eq:T}
T_\epsilon(H)=N_\epsilon(H) \left(t_{1} + t_{2} H\right),
\end{equation}
where $t_1$ denotes the fixed overhead of a single round (including communication and aggregation cost),  and $t_2$ denotes the execution time to perform a single update on the worker. Since $N_\epsilon(H)$ is a  decreasing function in $H$ there is  an optimal value $H^\star$ minimizing the execution time $T_\epsilon$ in a given setting. As we will demonstrate this optimal value is very sensitive to the specific infrastructure the algorithm is executed on.
While finding a general function from for $N_\epsilon(H)$ that can be used to model convergence for any dataset and/or algorithm remains an interesting research topic we will provide a model for  \cocoa in our experimental setup in Section \ref{sec:comcomp}.


\section{Programming Frameworks  for Distributed Computing}
\label{sec:tools}
There exist many different programming frameworks and libraries that are designed to simplify the implementation of distributed algorithms.
In this work we will focus on \spark, due to its widespread use, and compare it to the well established MPI framework. 

\vspace{-0.1cm}
\subsection{\spark} Apache \spark~\cite{Zaharia:2012RDD} is an open source  general-purpose cluster computing framework developed by the AMP lab at the University of California, Berkeley. The core concept underpinning \spark~is the resilient distributed dataset (RDD) which represents a read-only collection of data elements, spread across a  cluster that can be operated on in parallel. 
With this abstraction, \spark~allows the developer to describe their distributed program as a sequence of high level operations on RDDs  without being concerned with  scheduling, load-balancing and fault tolerance. 
The core of \spark~is written in Scala, runs on the Java virtual machine (JVM) and offers a functional programming API to Scala, Python, Java and R. The Python API is called \textit{py\spark}~and exposes the \spark~programming model to Python. 
Specifically, the local driver consists of a Python program in which a \spark~context is created. The \spark~context communicates with the Java virtual machine (over py4J) which in turn is responsible for initiating and communicating with Python processes. 

\vspace{-0.1cm}
\subsection{MPI}  Message Passing Interface (MPI) \cite{MPI93} is a  language-independent communication protocol for parallel computing that has been developed for high-performance computing (HPC) platforms. It offers a scalable and effective way of programming distributed systems consisting of up to tens of thousands of nodes. MPI allows application programmers to take advantage of problem-specific load-balancing, communication optimization techniques and various different ways of enabling fault-tolerance for distributed applications. However, this typically requires a significant amount of manual work and advanced understanding of the algorithms, MPI's library functions, and the underlying network architecture.


\section{Implementation Details}
To understand the characteristics of the aforementioned programming frameworks and their implications on the performance of distributed learning, we have chosen  the \cocoa algorithm \cite{CoCoA16} as a representative example and implemented it from scratch  on \spark, py\spark and MPI.
In our implementations these programming frameworks are used  to handle the communication of updates between workers during the training of \cocoa. Mathematically, all of the following implementations are equivalent but small differences in the learned model can occur due to randomization and slight variations in data partitioning, which needs to be implemented by the developer in the case of MPI. 
As an application we have chosen ridge regression because the least squares loss term, as well as the Euclidian norm regularizer,  
are separable functions, which allows us to apply and later compare all of the three algorithms mentioned in Section \ref{sec:alg}.

\subsection{Implementations}
\label{sec:Implementations}
\paragraph{($\mathrm{A}$) Spark} We use the open source implementation of Smith et al. \cite{SmithGit} 
as a reference implementation of \cocoa. This implementation is based on \spark~and entirely written in Scala. The Breeze library \cite{Breeze} is used to accelerate sparse linear algebra computations. As \spark does not allow for persistent local variables on the workers, the parameter vector $\alphav$ needs to be communicated to the master and back to the worker in every round, in addition to the shared vector $\vv$ -- the same applies to the following three \spark implementations.
\paragraph{($\mathrm{B}$) Spark+C} 
We replace the local solver of implementation \A~by a Java native interface (JNI) call to a compiled and optimized \cpp module. Furthermore, the RDD data structure is modified so that each partition consists of a flattened representation of the local data. 
In that manner, one can pass the local data into the native function call as pointers to contiguous memory regions rather than having to pass an iterator over a more complex data structure. The \cpp code is able to directly operate on the RDD data (with no copy) by making use of the \textit{GetPrimitiveArrayCritical} functions provided by the JNI. 
\paragraph{($\mathrm{C}$) pySpark} 
This implementation is equivalent to that of \A~except it is written entirely in Python/py\spark. The local solver makes use of the \textit{NumPy} package \cite{Numpy} for fast linear algebra. 
\paragraph{($\mathrm{D}$) pySpark+C} 
We replace the local solver of implementation \rC~with a function call to a compiled and optimized \cpp module, using the Python-C API. Unlike implementation \B~we did not flatten the RDD data structure since this was found to lead to worse performance in this case. 
Instead, we iterate over the RDD within the local solver in order to extract from each record a list of \textit{NumPy} arrays.
The list of \textit{NumPy} arrays is then passed into the \cpp module. The Python-C API allows \textit{NumPy} arrays to expose a pointer to their raw data and thus the need to copy data into any additional \cpp data structures is eliminated. 
\paragraph{($\mathrm{E}$) MPI}
The MPI implementation is entirely written in \cpp using the same code for the local solver used in \B~and \D. To initially partition the data we have developed  a custom load-balancing algorithm to distribute the computational load evenly across workers, such that  $\sum_{i\in \bP_k}\|\cv_i\|_0$ is roughly equal for each partition. 
This was found to perform comparable to the \spark~partitioning.


\vspace{-0.05cm}
\subsection{Infrastructure}
\label{sec:Infrastructure}

All our experiments are run on a cluster of 4 physical nodes interconnected in a LAN topology through a 10Gbit-per-port switched inter-connection.
Each node is equipped with 64GB DDR4 memory, an 8-core Intel Xeon$^*$
E5 x$86\_64$ 
CPU and  solid-state disks. 
The software configuration of the cluster is based on Linux$^*$ kernel v3.19, MPI v3.2, and Apache Spark v2.2. We use the Open MPI branch of MPI. Spark is configured to use 8 \spark~executors with 24 GB of memory each, 2 on each machine.
Furthermore, \spark does not use the HDFS filesystem; instead SMB sharing directly over ext4 filesystem I/O is employed. 
While this decision may occasionally lead to reduced performance in Spark, it
eliminates I/O measurement delay-variation artifacts 
which enables a fairer comparison with MPI since all overheads measured are strictly related to Spark. 



\section{analysis and optimization of \spark}
\label{sec:Results}
In the first part of this section we analyze the performance of the different implementations of the \cocoa algorithm discussed in Section~\ref{sec:Implementations}  by training the ridge regression model on the publicly available \textit{webspam} dataset \cite{webspam}.

\begin{figure}[b]
\vspace{-0.3cm}
\centering
%
%

%
\begin{tikzpicture}
\begin{axis}[%
width=2.7in, 
height=1in,
at={(0 in,0 in)},
scale only axis,
bar width=10,
separate axis lines,
every outer x axis line/.append style={black},
every x tick label/.append style={font=\color{black}},
every x tick/.append style={black},
xmin=0,
xmax=900,
xtick={0,100,200,300,400,500,600, 700,800,900},
xticklabels={{\small 0},{\small 100},{\small 200},{\small 300},{\small 400},{\small 500},{\small 600},{\small 700},{\small $\dots$},{\small 12000}},
xlabel={\small time [s]},
every outer y axis line/.append style={black},
every y tick label/.append style={font=\color{black}},
every y tick/.append style={black},
ymin=0.5,
ymax=5.5,
ytick={1,2,3,4,5},
yticklabels={{\E},{\D},{\rC},{\B},{\A}},
 yticklabel style={text width=0.8cm, align=left}, 
axis background/.style={fill=white},
xmajorgrids,
ymajorgrids,
legend style={legend columns = -1, legend cell align=left, align=left, draw=black,at={(0.98,0.04)},anchor=south east}
]
\addplot[xbar stacked, fill=white, draw=black,postaction={pattern=north west lines}] table[row sep=crcr] {%
0 1	\\
0 2	\\
0 3	\\
0 4	\\
0 5 \\
};
\addlegendentry{\small$T_{\text{master}}$}
\addplot[xbar stacked, fill=white, draw=black,postaction={pattern=north east lines}] table[row sep=crcr] {%
0 1	\\
0 2	\\
0 3	\\
0 4	\\
0 5 \\
};
\addlegendentry{\small$T_{\text{overhead}}$}
\addplot[xbar stacked, fill=white, draw=black,postaction={pattern=dots}] table[row sep=crcr] {%
0 1	\\
0 2	\\
0 3	\\
0 4	\\
0 5 \\
};
\addlegendentry{\small$T_{\text{worker}}$}


\addplot[xbar stacked, fill=blue, draw=black,postaction={pattern=north west lines}] table[row sep=crcr] {%
1.8	5\\
0	4\\
0	3\\
0	2\\
0	1\\
};
\addplot[xbar stacked, fill=blue, draw=black,postaction={pattern=north west lines}] table[row sep=crcr] {%
0 5\\
1.8	4\\
0	3\\
0	2\\
0	1\\
};
\addplot[xbar stacked, fill=red, draw=black,postaction={pattern=north west lines}] table[row sep=crcr] {%
0	5\\
0	4\\
1.7	3\\
0 2\\
0	1\\
};
\addplot[xbar stacked, fill=red, draw=black,postaction={pattern=north west lines}] table[row sep=crcr] {%
0	5\\
0	4\\
0	3\\
1.7	2\\
0	1\\
};
\addplot[xbar stacked, fill=green, draw=black,postaction={pattern=north west lines}] table[row sep=crcr] {%
0	5\\
0	4\\
0	3\\
0	2\\
0.01	1\\
};

\addplot[xbar stacked, fill=white!50!blue, draw=black,postaction={pattern=north east lines}] table[row sep=crcr] {%
143.2	3\\
0	2\\
0	5\\
0	4\\
0	1\\
};
\addplot[xbar stacked, fill=white!50!blue, draw=black,postaction={pattern=north east lines}] table[row sep=crcr] {%
0 3\\
186.5	2\\
0	5\\
0	4\\
0	1\\
};
\addplot[xbar stacked, fill=white!50!red, draw=black,postaction={pattern=north east lines}] table[row sep=crcr] {%
0	3\\
0	2\\
82.2	5\\
0	4\\
0	1\\
};
\addplot[xbar stacked, fill=white!50!red, draw=black,postaction={pattern=north east lines}] table[row sep=crcr] {%
0	3\\
0	2\\
0	5\\
20.8	4\\
0	1\\
};
\addplot[xbar stacked, fill=white!50!green, draw=black,postaction={pattern=north east lines}] table[row sep=crcr] {%
0	3\\
0	2\\
0	5\\
0	4\\
0.99	1\\
};

303763.0,389914

\addplot[xbar stacked, fill=blue, draw=black,postaction={pattern=dots}] table[row sep=crcr] {%
740	3\\
0	2\\
0	5\\
0	4\\
0	1\\
};

\addplot[xbar stacked, fill=blue, draw=black,postaction={pattern=dots}] table[row sep=crcr] {%
0	3\\
53.7	2\\
0	5\\
0	4\\
0	1\\
};
\addplot[xbar stacked, fill=red, draw=black,postaction={pattern=dots}] table[row sep=crcr] {%
0 3\\
0	2\\
310	5\\
0	4\\
0	1\\
};
\addplot[xbar stacked, fill=red, draw=black,postaction={pattern=dots}] table[row sep=crcr] {%
0	3\\
0	2\\
0	5\\
60.7	4\\
0	1\\
};
\addplot[xbar stacked, fill=green, draw=black,postaction={pattern=dots}] table[row sep=crcr] {%
0	3\\
0	2\\
0	5\\
0	4\\
42.7	1\\
};
\end{axis}
\end{tikzpicture}%
\caption{{Total run time for 100 iterations with $H=n_k$ split into compute time and overheads for the \spark implementations \A~and \B, the py\spark~implementations \rC~and \D~and the MPI implementation~\E.}}
\label{fig:barPlot-a}
\vspace{-0.15cm}
\end{figure}
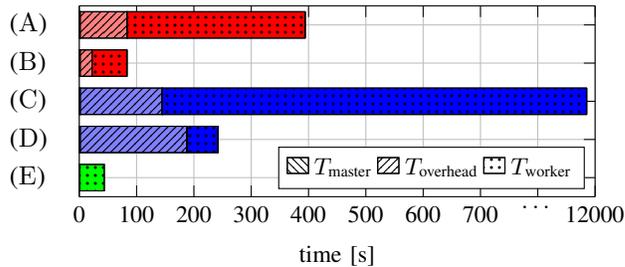

\subsection{\spark~overheads}
\label{sec:sparkOverheads}
We start by extracting the computational time from the total run time for the individual implementations. 
Therefore we fixed the number of rounds, as well as $H$, and measured, for each implementation, the total execution time ($T_\text{tot}$), as well as the time spent computing on each worker ($T_\text{worker}$) and the time spent computing on the master ($T_\text{master}$). We denote $T_\text{overhead} := T_\text{tot}-T_\text{worker}-T_\text{master}$
which measures overheads related to communications including data transfer as well as serialization/deserialization overheads. The results are displayed in Figure \ref{fig:barPlot-a}. 

We observe that the performance of the \spark \A~and py\spark \rC~implementations is vastly dominated by the time spent in the local solver. While the code written in Scala performs significantly better than the equivalent Python implementation, both can be accelerated significantly by replacing the local solver with \cpp~modules. 
Thereby, the local execution time of the \spark implementation is reduced by a factor of $6$ and the execution time of the py\spark implementation by more than $2$ orders of magnitude. The local execution time of the \cpp code is roughly the same for implementation \B, \D~and \E~up to some internal overheads of the JNI.
Leaving the language-dependent differences in execution time aside, focusing on the overheads and subtracting the actual communication cost, as measured in the MPI implementation ($3\%$ of the total execution time), we can accurately quantify the framework-related overheads of \spark~(and~py\spark). We can see that the overheads of the py\spark~implementation \rC~are $2\times$ larger than those of the reference \spark~implementation \A~written in Scala. This performance loss of py\spark was also observed in earlier work \cite{Holden} and can be attributed to the Python API which introduces additional serialization steps and overheads associated with initializing Python processes and copying data between the JVM and the Python interpreter.
Furthermore, we see that calling the \cpp modules from Python adds some additional overhead on top of the py\spark~overhead, which can be attributed to the large number of Python-C API calls that are required.
However, despite the slight increase, these additional overheads are negligible compared to the gain in execution time achieved by offloading the local solver into \cpp.
For Scala we do not see the same increase in overheads and, in fact, observe the opposite behavior.
We believe that this improvement can be attributed to the flattened RDD data format that was implemented when adding the \cpp modules in Scala.
This structure was explicitly designed to minimize the number of JNI calls. 
We can see that this flattened data format brings a large benefit: overheads are reduced by a factor of $3$. 
We have also implemented the flattened format in Python but we were not able to achieve a similar improvement.


\vspace{-0.1cm}
\subsection{Reducing \spark~Overheads}
\label{sec:hack}
Before we further analyze the implications of the overheads of the \spark framework on the achievable performance of \cocoa we will propose two techniques for extending the functionality of \spark~so that these overheads  can be somewhat alleviated for distributed learning algorithms.

\paragraph{ Persistent Local Memory} 
\spark~does not allow for local variables on the workers that can persist across stage boundaries, that is, the algorithm rounds. 
Thus the \cocoa algorithm, as well as mini-batch SCD, require additional communication when implemented in \spark since it is not possible for workers to store their dedicated coordinates of $\alphav$ locally. 
As a consequence, in addition to the shared vector, the $\alphav$ vectors need to be communicated to the master and back in every stage of the algorithm, thus increasing the overhead associated with communication.  
However, it is relatively straightforward to provide such functionality from within the \cpp extension modules. Globally-scoped arrays can be allocated upon first execution of the local solver that store the local $\alphav$ vectors. 
The state of these arrays persists into the next stage of execution in \spark, and thus the additional communication is no longer necessary. It should be noted that this extension comes at the expense of a violation of the \spark~programming model in terms of consistency of external memory with the lineage graph. 

\paragraph{ Meta-RDDs} For the Python implementations in particular, there is a significant overhead related to the RDD data structure. It is possible to overcome this overhead by following an approach similar to that in \cite{TensorFrames} and working with RDDs that consist purely of meta-data (e.g. feature indices) and handling all loading and storage of the training data from within underlying native functions. 
While some additional effort is required to ensure data resiliency, \spark~is still being used to schedule execution of the local workers, to collect and aggregate the local updates and broadcast the updated vectors back to the workers. 

\vspace{0.2cm}

\begin{figure}[t]
\centering
%
%

%
\begin{tikzpicture}
\begin{axis}[%
width=2.7in,
height=1in,
at={(0 in, 0 in)},
scale only axis,
bar width=10,
separate axis lines,
every outer x axis line/.append style={black},
every x tick label/.append style={font=\color{black}},
every x tick/.append style={black},
xmin=0,
xmax=250,
xtick={0,50,100,150,200,250},
xticklabels={{\small 0},{\small 50},{\small 100},{\small 150},{\small 200},{\small 250}},
xlabel={\small time [s]},
every outer y axis line/.append style={black},
every y tick label/.append style={font=\color{black}},
every y tick/.append style={black},
ymin=0.5,
ymax=5.5,
ytick={1,2,3,4,5},
yticklabels={{\E},{\D$^\star$},{\D},{\B$^\star$},{\B}},
yticklabel style={text width=0.8cm, align=left},
axis background/.style={fill=white},
xmajorgrids,
ymajorgrids,
legend style={legend columns = -1,legend cell align=left, align=left, draw=black,at={(0.98,0.04)},anchor=south east}
]
\addplot[xbar stacked, fill=white, draw=black,postaction={pattern=north west lines}] table[row sep=crcr] {%
0 1	\\
0 2	\\
0 3	\\
0 4	\\
0 5 \\
};
\addlegendentry{\small$T_{\text{master}}$}
\addplot[xbar stacked, fill=white, draw=black,postaction={pattern=north east lines}] table[row sep=crcr] {%
0 1	\\
0 2	\\
0 3	\\
0 4	\\
0 5 \\
};
\addlegendentry{\small$T_{\text{overhead}}$}
\addplot[xbar stacked, fill=white, draw=black,postaction={pattern=dots}] table[row sep=crcr] {%
0 1	\\
0 2	\\
0 3	\\
0 4	\\
0 5 \\
};
\addlegendentry{\small$T_{\text{worker}}$}


\addplot[xbar stacked, fill=blue, draw=black,postaction={pattern=north west lines}] table[row sep=crcr] {%
1.8	3\\
0	2\\
0	5\\
0	4\\
0	1\\
};
\addplot[xbar stacked, fill=blue, draw=black,postaction={pattern=north west lines}] table[row sep=crcr] {%
0 3\\
1	2\\
0	5\\
0	4\\
0	1\\
};
\addplot[xbar stacked, fill=red, draw=black,postaction={pattern=north west lines}] table[row sep=crcr] {%
0	3\\
0	2\\
1.7	5\\
0 4\\
0	1\\
};
\addplot[xbar stacked, fill=red, draw=black,postaction={pattern=north west lines}] table[row sep=crcr] {%
0	3\\
0	2\\
0	5\\
1	4\\
0	1\\
};
\addplot[xbar stacked, fill=green, draw=black,postaction={pattern=north west lines}] table[row sep=crcr] {%
0	3\\
0	2\\
0	5\\
0	4\\
0.01	1\\
};

\addplot[xbar stacked, fill=white!50!blue, draw=black,postaction={pattern=north east lines}] table[row sep=crcr] {%
186.5	3\\
0	2\\
0	5\\
0	4\\
0	1\\
};
\addplot[xbar stacked, fill=white!50!blue, draw=black,postaction={pattern=north east lines}] table[row sep=crcr] {%
0 3\\
18.21	2\\
0	5\\
0	4\\
0	1\\
};
\addplot[xbar stacked, fill=white!50!red, draw=black,postaction={pattern=north east lines}] table[row sep=crcr] {%
0	3\\
0	2\\
20.8	5\\
0	4\\
0	1\\
};
\addplot[xbar stacked, fill=white!50!red, draw=black,postaction={pattern=north east lines}] table[row sep=crcr] {%
0	3\\
0	2\\
0	5\\
6.5	4\\
0	1\\
};
\addplot[xbar stacked, fill=white!50!green, draw=black,postaction={pattern=north east lines}] table[row sep=crcr] {%
0	3\\
0	2\\
0	5\\
0	4\\
0.99	1\\
};

\addplot[xbar stacked, fill=blue, draw=black,postaction={pattern=dots}] table[row sep=crcr] {%
53.7	3\\
0	2\\
0	5\\
0	4\\
0	1\\
};

\addplot[xbar stacked, fill=blue, draw=black,postaction={pattern=dots}] table[row sep=crcr] {%
0	3\\
50.57	2\\
0	5\\
0	4\\
0	1\\
};
\addplot[xbar stacked, fill=red, draw=black,postaction={pattern=dots}] table[row sep=crcr] {%
0 3\\
0	2\\
60.7	5\\
0	4\\
0	1\\
};
\addplot[xbar stacked, fill=red, draw=black,postaction={pattern=dots}] table[row sep=crcr] {%
0	3\\
0	2\\
0	5\\
61.36	4\\
0	1\\
};
\addplot[xbar stacked, fill=green, draw=black,postaction={pattern=dots}] table[row sep=crcr] {%
0	3\\
0	2\\
0	5\\
0	4\\
42.7	1\\
};

\end{axis}
\end{tikzpicture}%
\vspace{-0.1cm}
\caption{The performance of the optimized implementations \B$^\star$ and \D$^\star$: by introducing persistent local variables and meta-RDDs we are able to significantly reduce the overheads of \spark relative to MPI.}
\label{fig:barPlot-b}
\vspace{-0.5cm}
\end{figure}
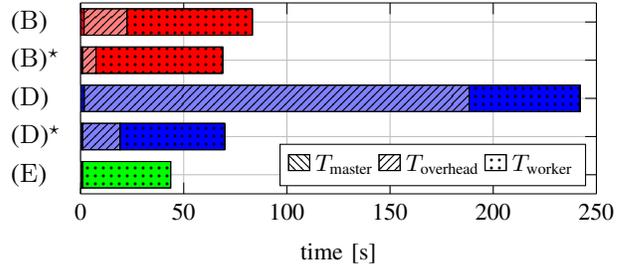

We have implemented these two features for both the Scala and the Python-based implementations of \cocoa. In Figure \ref{fig:barPlot-b} we compare the execution time and the overheads of these optimized implementations \B$^*$ and \D$^*$ with the corresponding implementations that only make use of native functions. We observe that our two modifications reduce overheads of the Scala implementation \B~by $3\times$ and those of the Python implementation \D~by $10\times$.
For the Scala implementation, the overall improvement due to using the meta-RDDs is small and most of the gain comes from introducing local memory and thus reducing the amount of data that needs to be communicated. However, for the Python implementation the effect of using meta-RDDs is far more significant. This is most likely due to the vast reduction in inter-process communication that has been achieved. 
It is worth pointing that the concept of meta-RDDs has additional applications and implications since similar techniques have been used to overcome some of the limitations of \spark, such as using GPUs inside \spark \cite{IBMGPUenabler}.

\begin{figure*}[t]
\centering
\subfloat[Time to achieve training suboptimality $10^{-3}$  for implementations \A-\E~as a function of $H$.]{\includegraphics[width=0.65 \columnwidth]{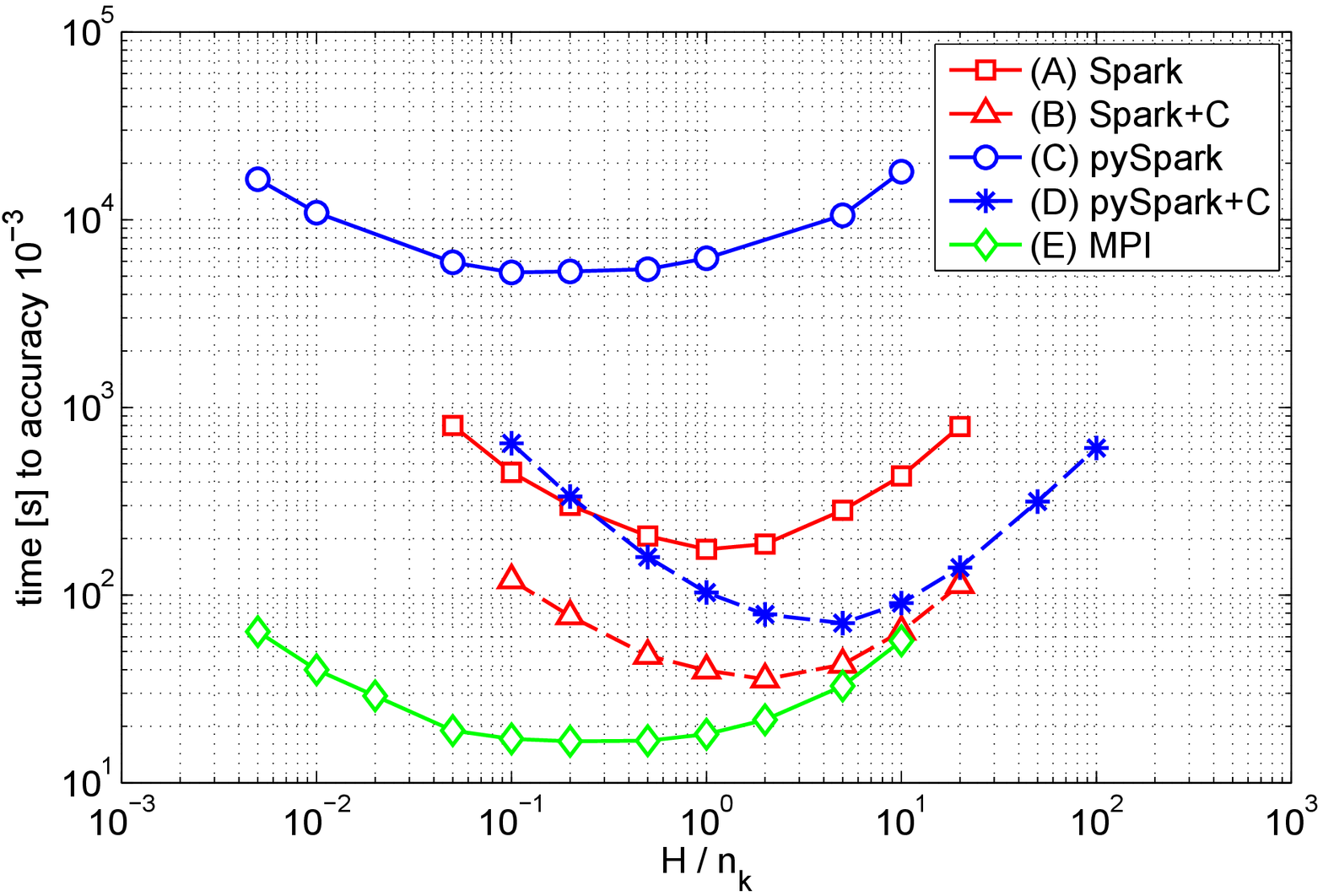}\label{fig:H-a}}
\hspace{0.5cm}
\subfloat[The optimal value of $H$ as a function the ratio $t_1/t_2$ of communication time and normalized computation time.]{\includegraphics[width=0.63\columnwidth]{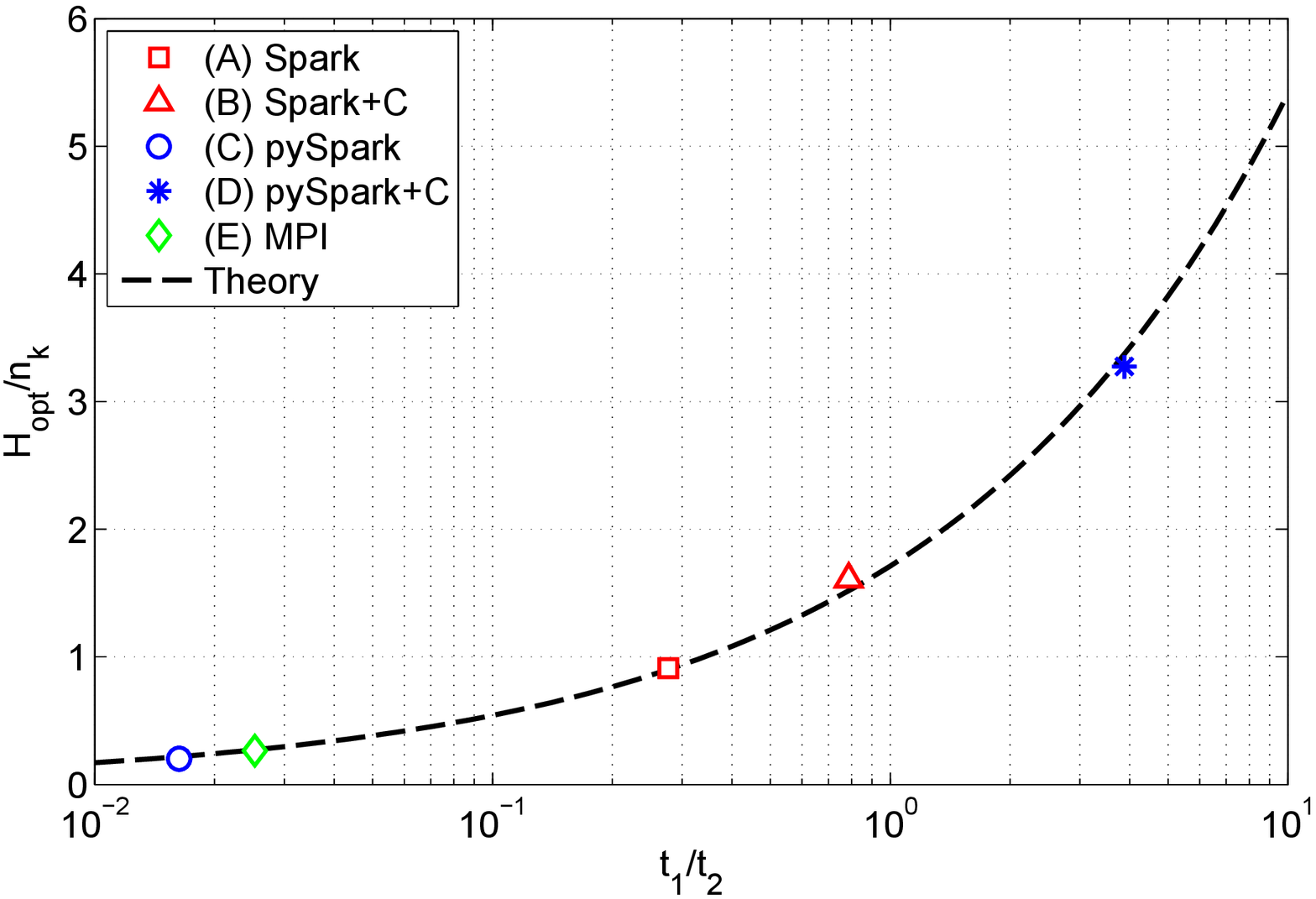}\label{fig:H-b}}
\hspace{0.5cm}
\subfloat[Evolution of suboptimality over time for the implementations \A-\E~using optimized values for $H$ in all cases.]{\includegraphics[width=0.65\columnwidth]{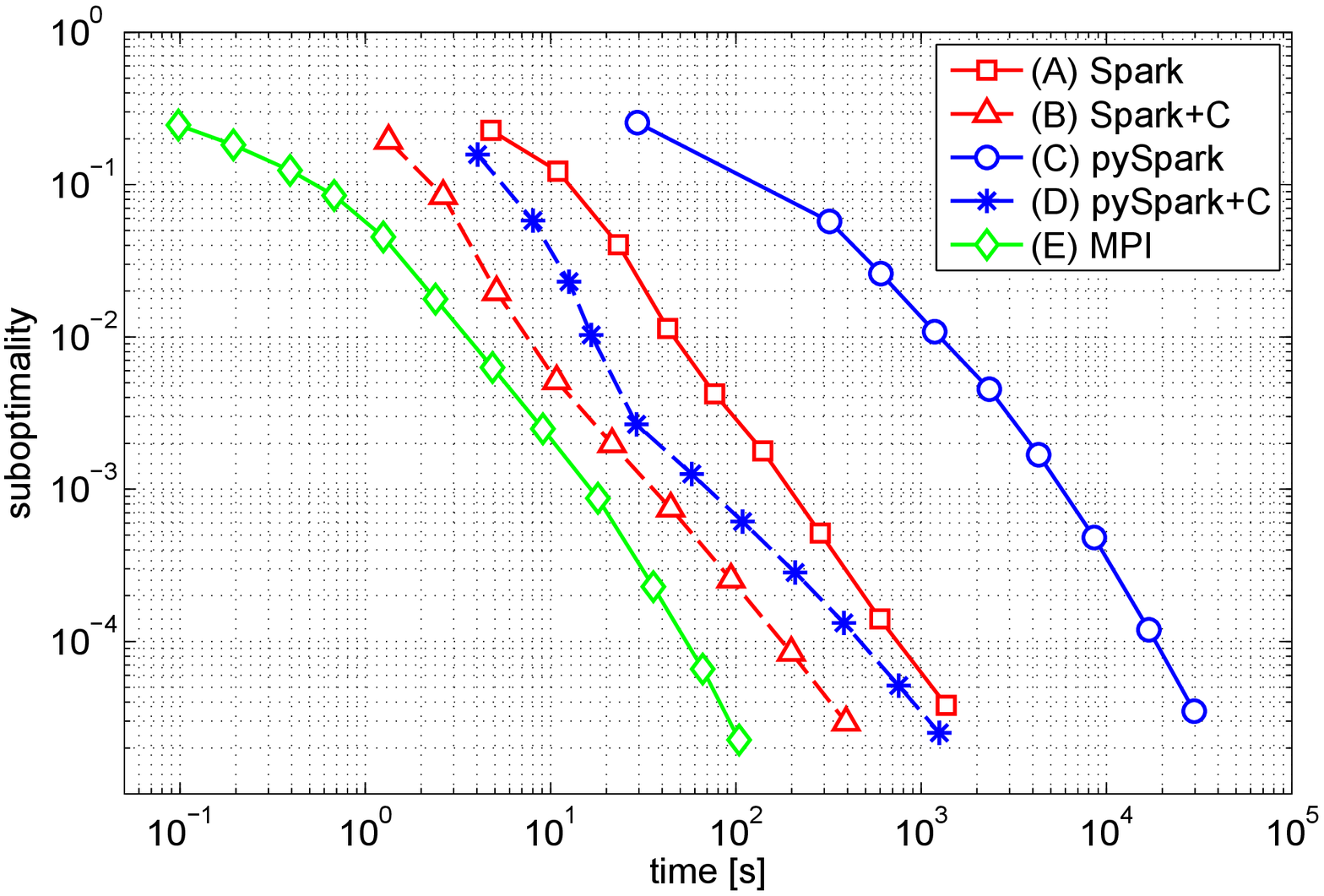}\label{fig:perf_opt}}
\caption{Optimization of the parameter $H$: trading-off communication vs. computation.}
\label{fig:H} 
\vspace{-0.5cm}
\end{figure*}

\section{Trading off Communication vs. Computation}
\label{sec:comcomp}

We  have seen in Figure \ref{fig:barPlot-a} that the implementations \A-\E~suffer from different computational efficiency and overheads associated with communication and data management. These are reflected by $t_2$ resp. $t_1$ in our model objective \eqref{eq:T}. Hence, to optimize performance, i.e., minimize $T_\epsilon$, for  the different implementations, it is essential to optimize the hyper-parameter $H$ separately for each implementation to account for these different costs of communication and computation. In this section we will study how the parameter $H$ can be used to control this trade-off for distributed implementations of machine learning and demonstrate the range of improvement an algorithm developer can expect when taking this approach.

In Figure \ref{fig:H-a} we show the time measured to achieve a suboptimality of $10^{-3}$ as a function of $H$ for the five different implementations of \cocoa \A-\E. 
We see that there is indeed an optimal trade-off for every implementation and the optimal value of $H$ varies significantly among the different implementations of the same algorithm on the same hardware. 
Hence, in order to get the best performance out of every implementation, it is crucial that $H$ be tuned carefully. 
Failure to do so may degrade performance dramatically. Indeed, we can see that the best performance of the py\spark implementation is achieved for $H=0.2 n_{k}$, i.e., every worker performs $0.2 n_{k}$ coordinate updates in every round. 
However, for the accelerated py\spark implementation \D~the optimal value of $H$ is more than $25\times$ larger. 
This is because, in implementation \D, the computational cost is significantly reduced relative to the vanilla py\spark implementation (see Figure \ref{fig:barPlot-a}) and we can afford to do more updates between two consecutive rounds of communication, thus obtaining a more accurate solution to the local subproblems. Hence, if  the algorithm was not adapted when  replacing the local solver by C++ modules, the gain observed would be only $15\times$ instead of $75\times$.
Also, comparing implementation \D~to the MPI implementation \E~for which the computation cost is the same but communication is much cheaper, we see a similar difference. While the overheads are less significant, the same reasoning applies to the Scala implementations. 
These results demonstrate that introducing a tunable hyper-parameter to trade-off communication and computation when designing a distributed learning algorithm is crucial for its applicability in practical environments.

\subsection{Theoretical Analysis}
To better understand this trade-off illustrated in Figure \ref{fig:H-a} we recall the performance model introduced in Section \ref{sec:model}. For the algorithm and dataset under discussion, $N_\epsilon(H)$ can accurately be modelled by:
 \begin{equation}
 \label{eq:N}
 N_\epsilon(H) =  \frac {a} H + b,
 \end{equation}
where $a,b\in\mathbb{R}$ are constants. Combining \eqref{eq:T} with \eqref{eq:N} and optimizing for $H$ yields
\[H_{opt} = c \sqrt{ {t_1}/{t_2}}\]
for some constant $c\in\mathbb{R}$.
Hence, for this algorithm and dataset, the optimal value $H$ is proportional to the square-root of the ratio between communication and computation cost, which could easily be measured as part of  a pre-training phase.  
In Figure \ref{fig:H-b} we show that this theoretical estimate precisely agrees with the measurements from Figure \ref{fig:H-a}.

\section{Performance Evaluation}

We start by comparing the performance of the five implementations \A-\E~presented in Section \ref{sec:Implementations} for individually optimized values of $H$, see Figure \ref{fig:perf_opt}. 
We observe that this optimization amplifies the performance differences observed in Figure \ref{fig:barPlot-a}. The comparison between implementation \B, \D~and \E~is of particular interest, because in these implementations the computations on the workers are unified in order to eliminate language dependent differences in computational efficiency. Hence, the gap in performance between \B~and \D~can solely be attributed to the overheads of using the Python API to \spark. Similarly, the performance difference between implementations \B, \D~and the MPI implementation \E~can be attributed to framework related overheads of \spark resp. py\spark over MPI. It is worth mentioning that when comparing \E~to \A~instead, more than half of the performance gap is due to the local solver computation being more efficient in \cpp than Scala.

\begin{figure*}[t]
\vspace{-0.45cm}
\centering
\captionsetup{width=0.9\columnwidth}%
\subfloat[Optimized implementations of the CoCoA algorithm to train the ridge regression model on the webspam dataset.]
{\includegraphics[width=7cm]{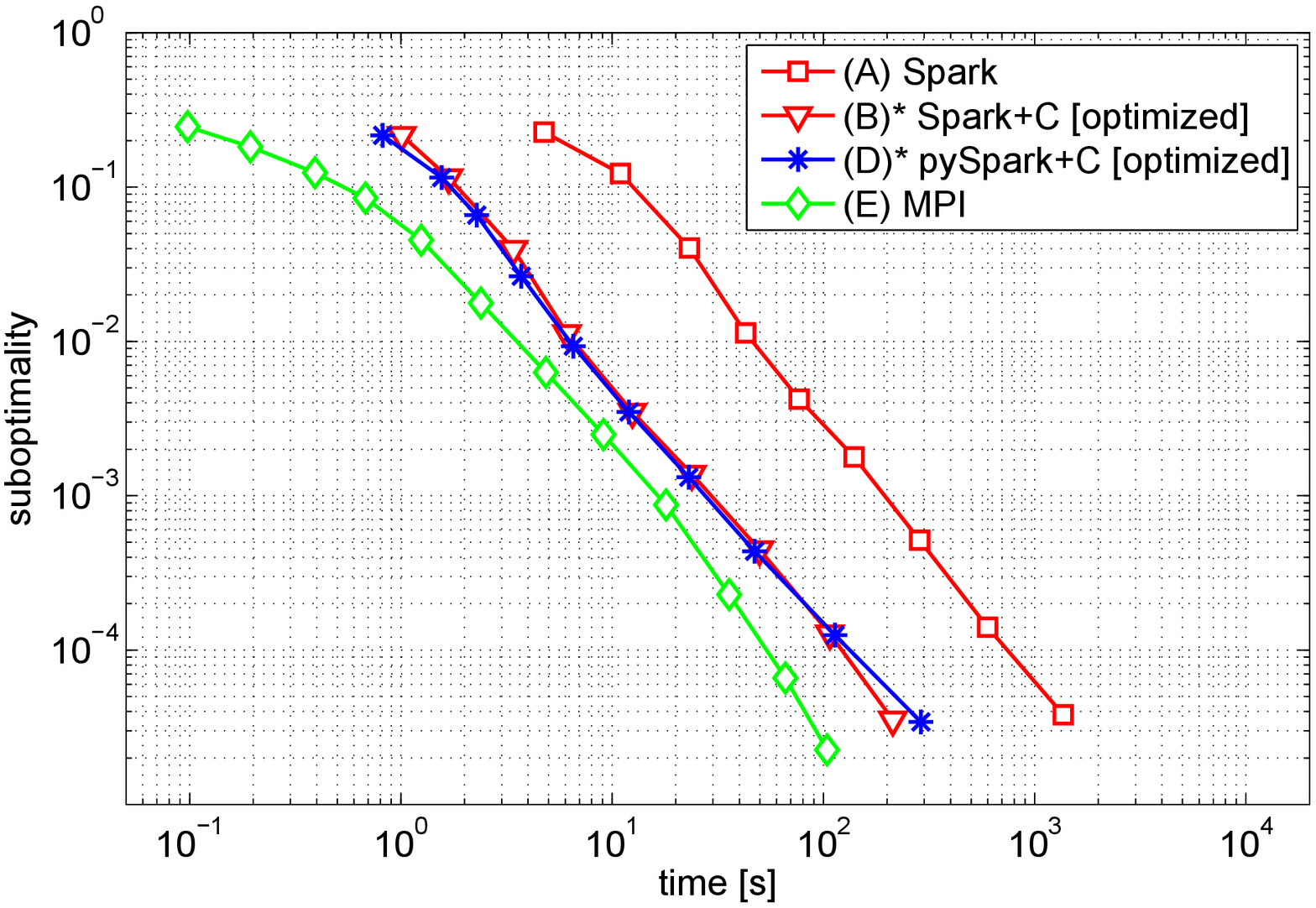}\label{fig:hack} }
\hspace{1.5 cm}
\subfloat[Training the ridge regression model on the webspam dataset using \cocoa, mini-batch SGD and mini-batch SCD. ]
{
%
%

\begin{tikzpicture}
\begin{axis}[%
width=2.5in,
height=1.6in,
at={(0in,0in)},
scale only axis,
bar width=10,
separate axis lines,
every outer x axis line/.append style={black},
every x tick label/.append style={font=\color{black}},
every x tick/.append style={black},
xmin=0,
xmax=3.8,
xtick={0.7,1.9,3.2},
xticklabels={{\footnotesize \cocoa},{\footnotesize mini-batch SCD},{\footnotesize mini-batch SGD}},
every outer y axis line/.append style={black},
every y tick label/.append style={font=\footnotesize },
every y tick/.append style={black},
ymin=1,
ymax=10000,
ymode =log,
ytick={1,10,100,1000,10000},
yticklabels={{$10^0$},{$10^1$},{$10^2$},{$10^3$},{$10^4$}},
y label style={at={(axis description cs:0.1,.5)},anchor=south},
ylabel={\footnotesize time [s] to suboptimality $10^{-2}$\vspace{-1cm}},
axis background/.style={fill=white},
xmajorgrids,
ymajorgrids,
legend style={
                    at={( 0.01,0.98)},
                    anchor=north west,
                    legend columns=-1,
                    /tikz/every even column/.append style={column sep=0.1cm}
                        }
]

\addplot[ybar stacked, fill=red, draw=black] table[row sep=crcr] {%
10	 45\\
12	1029\\
22	2000\\
};
\addlegendentry{\footnotesize {\A~Spark}~ };
\addplot[ybar stacked, fill=white!50!red, draw=black,postaction={pattern=north east lines}] table[row sep=crcr] {%
10	6\\
13	151\\
20	105\\
};
\addlegendentry{\footnotesize {\B$^\star$ {Spark}+C [opt.]~} };
\addplot[ybar stacked, fill=green, draw=black] table[row sep=crcr] {%
10	4\\
20	25\\
30	16\\
};
\addlegendentry{\footnotesize {\E~MPI}~};
\addplot[ybar, fill=red, draw=black] table[row sep=crcr] {%
0.4	 60\\
1.6	1029\\
2.8	2000\\
};
\addplot[ybar, fill=red!50!white, draw=black,postaction={pattern=north east lines}] table[row sep=crcr] {%
0.7	6\\
1.9	151\\
3.1	105\\
};
\addplot[ybar, fill=green, draw=black] table[row sep=crcr] {%
1.0	4\\
2.2	25\\
3.4	16\\
};

\end{axis}
\end{tikzpicture}%
\label{fig:others} }

\captionsetup{width=2\columnwidth}%
\caption{Proposed Spark optimizations: Performance results}
\vspace{-0.5cm}
\end{figure*}

By implementing the extensions suggested in Section \ref{sec:Implementations} in addition to the \cpp modules we managed to further improve the \spark performance by $25\%$ and the py\spark performance by $63\%$.
This is shown in Figure \ref{fig:hack}. 
We would like to emphasize that while reducing overheads improves performance by reducing the absolute time spent communicating, it provides the additional benefit that the value of $H$ can be reduced and thus communication frequency is increased, resulting in faster convergence of the algorithm. 
Without being offered the possibility of tuning $H$ we would only be able to achieve $50\%$ of the performance gain observed by implementing our extensions.
Thus, by combining our optimizations we can reduce the performance gap between \spark, resp. py\spark, and MPI from $10\times$, resp. $20\times$ to an acceptable level of less than $2\times$. 
While we acknowledge that this performance improvement has come at the expense of implementation complexity, these extensions could be integrated within a new or existing \spark library and thus effectively hidden from the developer building a larger machine learning pipeline.

The proposed techniques to improve the performance of distributed machine learning executed on frameworks such as \spark do not only apply to \cocoa. 
In fact, these techniques can be useful for any algorithm fitting the synchronous pattern of communication  described in Figure \ref{fig:cluster}.
To illustrate this, we have implemented mini-batch SGD and mini-batch SCD using the proposed optimizations and, in Figure \ref{fig:others}, compare the performance to a reference \spark implementation of both algorithms 
as well as the \cocoa implementations \A, \B$^\star$ and \E~that have been previously examined. 
To implement mini-batch SCD, we modified the local solver of the \cocoa implementations so that a mini-batch coordinate update is computed in each round.
For mini-batch SGD, we modified the data partitioning to distribute the data by samples, and used as a reference the implementation provided by \spark MLlib. Optimized \spark and MPI implementations were also developed.  The stepsize has been carefully tuned for mini-batch SGD and mini-batch SCD.
We observe that \cocoa performs better than the two other algorithms, which is consistent with the results in \cite{CoCoA16}. 
The gain from our proposed improvements to \spark is significant for all three algorithms, but for \cocoa we get significantly closer to the performance of MPI than for mini-batch SCD and mini-batch SGD. 
This is because the two other algorithms have different convergence properties (captured by $N_\epsilon(H)$) and require more frequent communication to achieve convergence. Hence, the overheads associated with communication -- which are larger for \spark than for MPI -- have a bigger effect on performance.

Finally, we evaluate the performance for the best of the three algorithms (\cocoa) across a range of different datasets. In Table \ref{tab:datasets} we present the training time for the \spark reference implementation \A, our optimized implementation \B$^\star$ and the MPI implementation \E~for five different datasets.
We could not run the reference \spark code for the kdda dataset on our cluster because it ran out of memory due to the large number of features.
We observe that by using the proposed optimizations the average performance loss of Spark relative to MPI has been reduced from approximately $20\times$ to around $2\times$.

\begin{table*}
  \caption{\spark optimizations of \cocoa for different datasets}
  \label{tab:datasets}
  \centering
  \begin{tabular}{crrcccccccc}
    \toprule
    Dataset & \# samples & \# nonzero features && \multicolumn{3}{c}{time  [s] to reach suboptimality $10^{-3}$}& &  \multicolumn{2}{c}{slow-down vs. MPI} \\
       &  &  && \spark & \spark optimized & MPI&&  \spark  & \spark optimized\\
    \midrule
  news20-binary&	19996	&	1355191&&		29.92&	2.26	&0.70 && 42.73 & 3.23\\
  webspam	&262938	&	680715&&		205.24&	29.11&	16.39 && 12.52 & 1.78\\
  E2006-log1p	&16087		&4265669	&&610.444&	83.04&	66.06 && 9.24 & 1.26\\
  url&	2396130		&3230442&&	1582.78	&216.96	&118.22 && 13.39 &	1.84\\
  kdda	&8407752		&19306083&&--	&184.51	&79.68&&--& 2.32\\
    \bottomrule
  \end{tabular}
  \vspace{-0.1cm}
\end{table*}

\section{Related Work}
There have been a number of previous efforts to study the performance of \spark and its associated overheads. 
In \cite{Reyes2015} a study was performed comparing the performance of large-scale matrix factorization in \spark and MPI. 
It was found that overheads associated with scheduling delay, stragglers, serialization and deserialization dominate the runtime in \spark, leading to significantly worse performance relative to MPI. 
The performance of \spark was also studied in \cite{DBLP:journals} for a number of data analytics benchmarks and it was found that time spent on the CPU was the bottleneck and the effect of improved network performance was minimal. The difference in performance between \spark and MPI/OpenMP was further examined in \cite{188988} for the k-nearest neighbors algorithm and support vector machines; the authors concluded that MPI/OpenMP outperforms \spark by over an order of 
magnitude.
Our work differs from these previous studies \cite{Reyes2015,DBLP:journals,188988} in that they consider a fixed  algorithm running on different frameworks, whereas we optimize the algorithm to achieve optimal performance for any specific framework and implementation, which we have demonstrated to be crucial for a fair analysis of machine learning workloads.

An approach to address \spark's computational bottlenecks, in a similar spirt to our extensions suggested in Section \ref{sec:hack}. was proposed in \cite{crotty2015tupleware}. The authors suggest a high-performance analytics system which they call Tupleware. Tupleware focus on improving the computation bottleneck of analytics tasks by automatically compiling user-defined function (UDF) centric workflows. In this context, a detailed comparison to \spark is provided in a single node setting, demonstrating the inefficiencies introduced by high-level abstractions like Java and iterators. While they suggest a novel analytics platform, our extensions aim to improve the performance of algorithms within a given framework.

The fundamental trade-off between communication and computation of parallel/distributed algorithms has been widely studied. It is well known that there is a fundamental limit to the degree of parallelization, after which adding nodes slows down performance. In the context of large-scale machine learning this behavior has been modelled in \cite{sparks2015automating} aiming to predict a reasonable cluster size. While such a model assumes increasing framework and communication overheads with the number of nodes in a cluster, their assumptions about algorithmic behavior are not reflective of the properties of iterative distributed algorithms, where convergence strongly depends on the communication frequency.

\section{Conclusions}
\label{sec:Conclusion}

In this work we have demonstrated that vanilla \spark implementations of distributed machine learning can exhibit a performance loss of more than an order of magnitude relative to equivalent implementations in MPI. A large fraction of this loss is due to language dependent overheads. After eliminating these overheads by offloading critical computations into \cpp, combining this with a set of practical extensions to \spark  and effective tuning of the algorithm, we demonstrated a reduction in this discrepancy with MPI to only $2\times$.
We conclude that in order to develop high-performance, distributed machine learning applications in Spark as well as other distributed computing frameworks, it is not enough to optimize the computational efficiency of the implementation. One must also carefully adapt the algorithm to account for the properties of the specific system on which such an application will be deployed. 
For this reason, algorithms that offer the user a tuning parameter to adapt to changes in system-level conditions are of considerable interest from a research perspective.

\section*{Acknowledgment}
The authors would like to thank Frederick R. Reiss from the IBM Spark Technology Center for highly constructive advice regarding this work and Martin Jaggi from EPFL for his help and inspiring discussions.

\vspace{0.3cm}

\footnotesize{$^*$ Intel Xeon is a trademarks or registered trademarks of Intel Corporation or its subsidiaries in the United States and other countries. Linux is a registered trademark of Linus Torvalds in the United States, other countries, or both.}
\vspace{-0.2cm}

\small
\bibliographystyle{IEEEtran}
\bibliography{bibnew,IEEEabrv}

\end{document}